\documentclass[conference]{IEEEtran}
\IEEEoverridecommandlockouts
\usepackage{url}
\usepackage[symbol,bottom]{footmisc}
\usepackage{multirow}
\usepackage{graphicx}
\usepackage[table,xcdraw]{xcolor}
\usepackage{cite}
\usepackage{tabularx}
\usepackage{booktabs} 
\usepackage{amsmath,amssymb,amsfonts}
\usepackage{algorithmic}
\usepackage{graphicx}
\usepackage{textcomp}
\usepackage{xcolor}
\def\BibTeX{{\rm B\kern-.05em{\sc i\kern-.025em b}\kern-.08em
    T\kern-.1667em\lower.7ex\hbox{E}\kern-.125emX}}
\begin{document}

\title{Attention-Based Beamformer For Multi-Channel Speech Enhancement}


\author{
    Jinglin Bai\textsuperscript{1,*}, Hao Li\textsuperscript{2,*}, Xueliang Zhang\textsuperscript{1}, Fei Chen\textsuperscript{2} \\
    \textsuperscript{1}College of Computer Science, Inner Mongolia University, Hohhot, China \\
    \textsuperscript{2}Department of Electrical and Electronic Engineering, \\Southern University of Science and Technology, Shenzhen, China \\
    bjlin@mail.imu.edu.cn, lih9@sustech.edu.cn, cszxl@imu.edu.cn, fchen@sustech.edu.cn
}

\maketitle

\footnotetext{* These authors contributed equally to this work.}
\begin{abstract}
Minimum Variance Distortionless Response (MVDR) is a classical adaptive beamformer that theoretically ensures the distortionless transmission of signals in the target direction, which makes it popular in real applications. Its noise reduction performance actually depends on the accuracy of the noise and speech spatial covariance matrices (SCMs) estimation. Time-frequency masks are often used to compute these SCMs. However, most mask-based beamforming methods typically assume that the sources are stationary, ignoring the case of moving sources, which leads to performance degradation. In this paper, we propose an attention-based mechanism to calculate the speech and noise SCMs and then apply MVDR to obtain the enhanced speech. To fully incorporate spatial information, the inplace convolution operator and frequency-independent LSTM are applied to facilitate SCMs estimation. The model is optimized in an end-to-end manner. Experiments demonstrate that the proposed method outperforms baselines with reduced computation and fewer parameters under various conditions. 
\end{abstract}


\section{Introduction}
\label{sec:introduction}
For microphone array signal processing [\citenum{benesty2008microphone}, \citenum{arrays2001signal}], most speech enhancement methods are based on fixed and adaptive beamformer techniques [\citenum{Benesty2012DifferentialMicrophone}, \citenum{Veen1988Beamforming}], leveraging the spatial characteristics of multi-channel data. Minimum variance distortionless response (MVDR) [\citenum{Veen1988Beamforming}] uses spatial information to calculate beamforming coefficients, ensuring that signals from the desired direction are transmitted without distortion.

Recent advancements in mask-based beamforming approaches [\citenum{heymann2016neural}, \citenum{erdogan2016improved}, \citenum{higuchi2016robust}, \citenum{kubo2019mask}] have attracted increased attention. The mask-based beamformer utilizes time-frequency masks derived from neural networks [\citenum{heymann2016neural}, \citenum{erdogan2016improved}] or other models [\citenum{higuchi2016robust}, \citenum{kubo2019mask}] to compute the spatial covariance matrices (SCMs), which capture the spatial information of the sources. These techniques are adaptable to different scenarios. When the sources remain stationary, SCMs are typically computed through temporal averaging. In scenarios involving moving sources, a dynamic approach is often employed to adapt to real-time variations, such as online [\citenum{higuchi2016robust}, \citenum{8683184}] or blockwise processing \cite{kubo2019mask}. However, the  dynamic adaptation faces the critical challenge of selecting an appropriate forgetting factor for online processing or determining the optimal block size for blockwise processing. 


ATT-MVDR \cite{ochiai2023mask} proposes a two-stage approach to address the problem. Initially, a neural network estimates the mask with the fifth channel as the target. Subsequently, two Transformer \cite{vaswani2017attention} encoders estimate the attention weights for the speech and noise instantaneous SCMs (ISCMs) \cite{ochiai2023mask} estimated by mask, respectively. However, ATT-MVDR is trained in two separate stages, which causes that the prediction of previous stage may include errors, such as information loss. These errors are then propagated to the second stage, leading to the  accumulation of errors. Moreover, 
 the traditional beamforming shows that the spatial information exists in each frequency bin of the array signal spectrum \cite{liu2010wideband}. Wideband beamforming is also independently processed in each frequency bin. During the estimation of attention weights by Transformer encoders, the frequency dimension is reduced or amplified through the linear layer, complicating the learning process of spatial information. Additionally, Transformer encoders have large parameters and high computational cost.

Recently, the Inplace Gated Convolutional Recurrent Neural Network (IGCRN) \cite{Liu2021Inplace} is proposed to learn spatial information across channels by discarding the down-sampling operation in the frequency dimension from the Convolutional Recurrent Network (CRN) \cite{Tan2019RealtimeSE}\cite{Tan2019LearningCS}. Due to the inplace characteristic of IGCRN, the spatial cues are explicitly maintained within each frequency bin. In addition, the LSTM layer is designed to capture temporal correlations across channels for each frequency bin, referred to as the frequency-independent LSTM. The IGCRN has been successfully applied to mono acoustic echo cancellation (AEC)\cite{Zhang2022Complex}, and stereo AEC\cite{Zhang2022LCSM2}.

\begin{figure*}[ht]
    \centering
    \includegraphics[width=0.945\linewidth, trim={1.3cm 1.40cm 1cm 0.95cm}, clip]{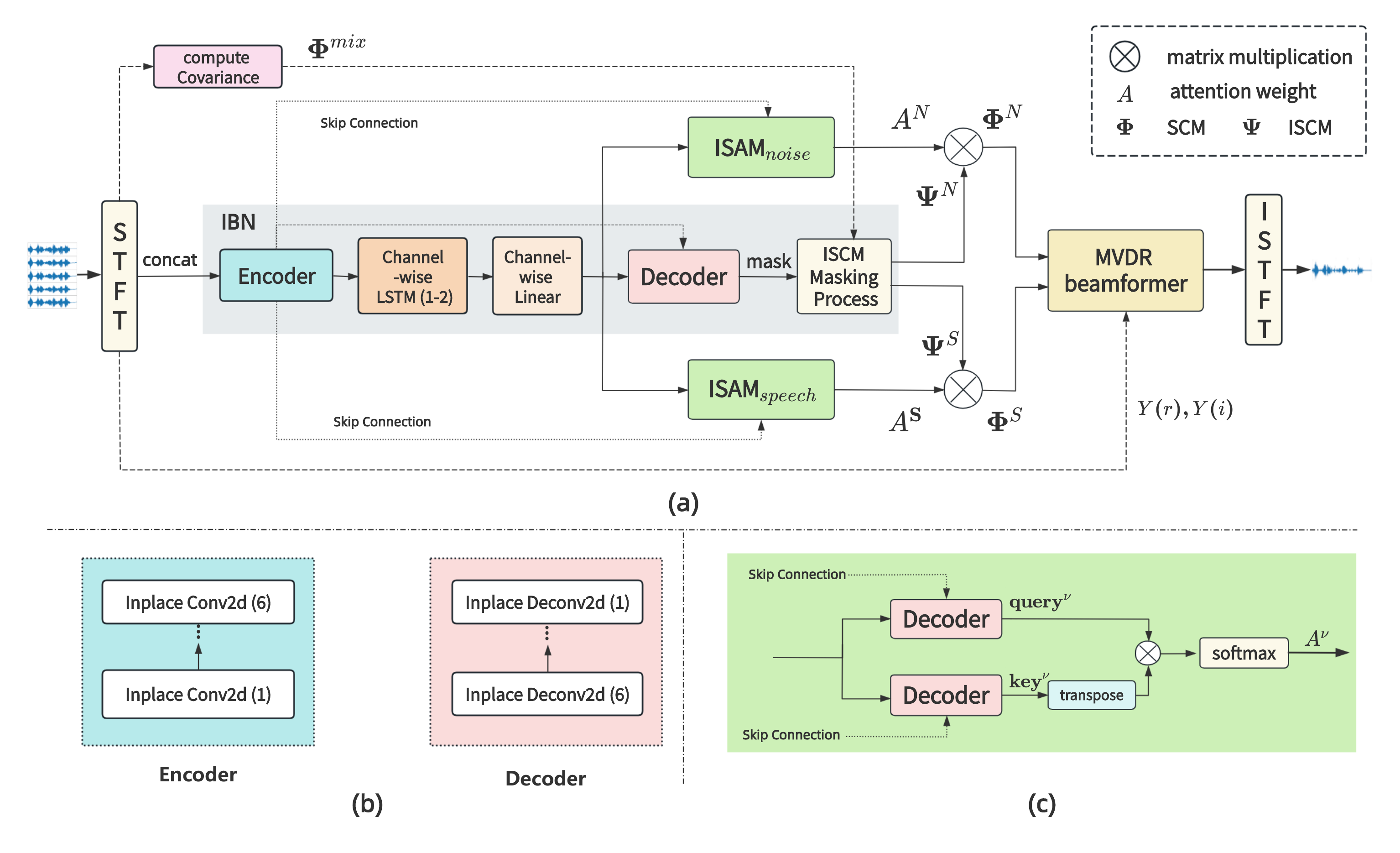}
    \caption{(a) An overview of ABIC-MVDR. (b) Structure of encoder and decoder. (c) Inplace self-attention module (ISAM).}
    \label{fig:overview}
\end{figure*}


In this paper, we propose a novel attention-based approach combined with the IGCRN and beamformer for multi-channel speech enhancement, referred to as ABIC-MVDR. The inplace convolution operator and two-layer frequency-independent LSTM are employed to learn spatial information  efficiently.  To better adapt to different scenarios, ABIC-MVDR utilizes the two decoder outputs of IGCRN to estimate temporal attention weights for the SCMs reconstruction, similar to the self-attention mechanism. By optimizing the model in an end-to-end manner, the accumulation of errors due to multiple stages can be avoided.
The main contributions of the proposed method are as follows: 
\begin{enumerate}
    \item The proposed attention-based beamformer precisely enhances the target, especially in dynamic scenarios.
    \item The IGCRN has been demonstrated to facilitate SCM estimation, enabling compatibility with beamforming.
    \item Our method achieves better performance than baselines, with lower computational costs and fewer parameters.
\end{enumerate}




\section{Methodology}

\subsection{Problem definition}
\( S^{m}_{t,f} \) and \( N^{m}_{t,f} \) denote the clean speech and background noise received by microphone \( m \) 
 , respectively. Let \( t \) and \( f \) represent the time frame index and frequency bin index, respectively. The received signal \( Y^{m}_{t,f} \) can be modeled as: 
\begin{align}
Y^{m}_{t,f} &= S^{m}_{t,f} + N^{m}_{t,f} \tag{1} \\
          &=  H^{m}_{t,f} S_{t,f}+ N^{m}_{t,f} \nonumber\; ,
\end{align}
where \( H^{m}_{t,f} \) indicates the acoustic impulse response from the target speaker to the \( m \)-th microphone.

\subsection{An overview of ABIC-MVDR}
ABIC-MVDR comprises three primary components: the IGCRN backbone network (IBN), the Inplace self-attention module (ISAM), and the MVDR beamformer. As shown in Fig.~\ref{fig:overview} (a), the signal collected by multiple microphones is processed through  Short-Time Fourier Transform (STFT). The real and imaginary parts are concatenated with the channel dimension and then fed into the encoder. The IBN is adopted to predict masks for the speech and noise ISCMs.  Two ISAMs are then utilized to allocate weights over time for the speech and noise SCMs reconstruction. Finally, the MVDR beamformer is applied to reconstruct the target speech.

Our proposed method comprises one encoder and five decoders. Two decoders are responsible for building the self-attention mechanism in the speech ISAM, while another two are used for building the self-attention mechanism in the noise ISAM. The remaining decoder is dedicated to mask prediction, which is utilized to construct the ISCM.

\subsection{ IGCRN backbone network (IBN)}\label{sec:IBN}
The encoder and decoder are shown in Fig.~\ref{fig:overview} (b). Inplace convolution refers to a convolutional neural network where the stride of the kernel is set to one, ensuring that features are not downsampled in the frequency dimension. This approach naturally and explicitly preserves spatial correlations within each frequency bin. Skip connections are used to concatenate the output of each Inplace Conv2d layer to the input of the corresponding Inplace Deconv2d layer. Following each Inplace Conv2d and Inplace Deconv2d layer, a batch normalization\cite{Ioffe2015BatchNorm} and an exponential linear unit (ELU) activation function are sequentially applied. A two-layer LSTM is utilized to model the temporal characteristics of each frequency bin across channels. Since the time delay for a specific direction is consistent across different frequency bins, they share the same LSTM units. Following this, a channel-wise linear layer is implemented to ensure that the output shape of the LSTM matches the input.

\subsection{Inplace Self-Attention Module (ISAM)}\label{sec:ISAM}
Unlike the transformer encoder architecture used in \cite{ochiai2023mask}, we estimate attention weights for the ISCM through a transposed matrix multiplication between the outputs from two IGCRN decoders in ISAM. In Fig.~\ref{fig:overview} (a), two ISAMs are utilized to allocate weights over time for reconstructing the speech SCM \(\Phi^S_{t,f}\) and the noise SCM  \(\Phi^N_{t,f}\), respectively.  As shown in Fig.~\ref{fig:overview} (c), the two decoders in ISAM$_{speech}$  or ISAM$_{noise}$ predict the $\mathbf{query^{\nu}}$ and $\mathbf{key^{\nu}}$ vectors, respectively. $\nu \in \{S, N\}$ are the indexs for speech and noise, respectively. The decoder structure in ISAM is similar to that described in \ref{sec:IBN}. And the $\tanh$ activation function is employed in the final layer.

After obtaining $\mathbf{query^{\nu}} \in \mathbb{R}^{F \times T \times D}$ and $\mathbf{key^{\nu}} \in \mathbb{R}^{F \times T \times D}$, the attention weights $\mathbf{A}^{\nu} \in \mathbb{R}^{F \times T \times T}$ are calculated by (2), where the softmax is applied on the second $T$ dimension. $T$ and $F$ represent the time frame index and frequency bin index, respectively, and $D$ represents the feature dimension. 

\begin{equation}
\mathbf{A}^{\nu} = \text{softmax}\left(\text{MASK}\left(\frac{\mathbf{query^{\nu}}(\mathbf{key^{\nu}})^T}{\sqrt{D}}\right)\right)\tag{2}\; .
\end{equation}

The MASK is a lower triangular matrix to ensure the self-attention mechanism is causal, as in  (3). Here, \( f \in \{0, 1, \dots, F-1\} \), and \( i, j \in \{0, 1, \dots, T-1\} \). If the model is operating under non-causal conditions, the MASK should be omitted.

\begin{equation}
\text{MASK}(x (f,i,j))= 
\begin{cases} 
x(f,i,j), & \text{if } i \geq j \\
-\infty, & \text{otherwise}
\end{cases}\tag{3}\; .
\end{equation}

The SCM $\boldsymbol{\Phi}^{\nu}_{t,f} \in \mathbb{C}^{M \times M}$ with $M$ channels at time-frequency bin $(t, f)$ is calculated as:


\begin{equation}
\boldsymbol{\Phi}^{\nu}_{t,f} = \mathbf{A}^{\nu}\mathbf{\Psi}^{\nu}_{t,f} \text{,} \tag{4}
\end{equation}
where the $\mathbf{\Psi}^{\nu}_{t,f} \in \mathbb{C}^{M \times M}$ refers to the estimated  ISCM obtained after the ISCM Masking Process and $ \mathbf{A}^{\nu}$ represents attention weights for ISCM  $\mathbf{\Psi}^{\nu}
_{t,f}$, as calculated in (2). 


\subsection{Minimum Variance Distortionless Response Beamformer}
In this paper, the beamforming filter coefficients  \(\mathbf{w}_{t,f} \in \mathbb{C}^M\) \cite{souden2010optimal} are calculated using MVDR, as shown in  (5). The $\boldsymbol{\Phi}^S_{t,f} \in \mathbb{C}^{M \times M}$ and $\boldsymbol{\Phi}^N_{t,f} \in \mathbb{C}^{M \times M}$ are the estimated SCMs of the speech and noise signals with $M$ channels at time-frequency bin $(t, f)$, respectively.
 The \(\mathbf{u} \in \mathbb{C}^M\) represents the index of the reference microphone.

\begin{equation}
\begin{split}
\mathbf{w}_{t,f} = \frac{(\boldsymbol{\Phi}^N_{t,f})^{-1} \boldsymbol{\Phi}^S_{t,f}}{\mathrm{Tr}((\boldsymbol{\Phi}^N_{t,f})^{-1} \boldsymbol{\Phi}^S_{t,f})} \mathbf{u}.
\end{split}
\tag{5}
\end{equation}

 The enhanced speech \(\hat{\mathbf{S}}_{t,f}\) is calculated through the observed signal \(\mathbf{Y}_{t,f}\), as shown in (6). And \((\cdot)^H\) represents the conjugate transpose.

\begin{equation}
\begin{split}
\hat{\mathbf{S}}_{t,f} = \mathbf{w}^H_{t,f} \mathbf{Y}_{t,f}.
\end{split}
\tag{6}
\end{equation}

\section{Experiments}

\subsection{Dataset and setting}
 We generate the scenario of moving speakers in noisy conditions for training and testing. The speech signals are acquired from the WSJ0 corpus\cite{Paul1992CSR}, while the noise signals are obtained from the CHiME-3 corpus\cite{Barker2015CHiME}. We design the experimental scenario to match that of \cite{ochiai2023mask}, with the only difference being that speech and noise signals are randomly paired at different signal-to-noise ratios (SNRs) ranging from -10 dB to 10 dB. After statistical analysis of the generated moving source signals, the minimum, maximum, and average source velocities \([m/s]\) (moving distance \([m]\)/utterance duration \([s]\)) are \(0.012\ m/s\), \(2.404\ m/s\), and \(0.238\ m/s\), respectively. We also create a non-moving dataset to evaluate scenarios with fixed-position speakers. Its configuration is identical to that of the moving dataset, except that the source speaker's position is fixed at the starting location. Through the above approach, we produce 30,000 noisy speech samples for the training set and 2,000 for the development and evaluation sets in both the non-moving and moving datasets, respectively.

\subsection{Model configuration}
The sampling frequency is configured to 16 kHz. Frame windowing is performed with a length of 320 samples and a stride of 160 samples. The Adam optimizer \cite{kingma2014adam} is applied during training, with a learning rate of \(1 \times 10^{-3}\). The batch size is set to 8.  The first channel is used as a reference. The model trains for 100 epochs guided by the SNR loss function \cite{LeRoux2019}. 
As shown in Table \ref{table0}, the size of each Inplace Conv2d and Inplace Deconv2d layer is specified in the [B, C, F ,T] format which is short for [\textit{Batchsize}, \textit{Channels}, \textit{Frequencies}, \textit{Time}]. The hyperparameters setting is given in the (\textit{kernelsize}, \textit{strides}, \textit{outchannels}) format for convolution, and (\textit{hidden\_layer\_size}) for channel-wise LSTM and Linear layers.

\begin{table}[th]
\centering
\setlength{\tabcolsep}{4pt} 
\renewcommand{\arraystretch}{0.96} 
\caption{The proposed ABIC-MVDR architecture.}
 \resizebox{\linewidth}{!}{%
 \begin{tabular}{|l|c|c|c|}
 \hline
 \textbf{Layer name} & \textbf{Input size} & \textbf{Hyperparameters} & \textbf{Output size} \\
 \hline
 Inplace Conv2d (1) & [B, 10, F, T] & 5$\times$1, (1, 1), 24 & [B, 24, F, T] \\
 \hline
 Inplace Conv2d (2-6) & [B, 24, F, T] & 5$\times$1, (1, 1), 24 & [B, 24, F, T] \\
 \hline
Reshape & [B, 24, F, T] & - & [B$\times$F, T, 24] \\
\hline
Channel-wise LSTM$\times$2 & [B$\times$F, T, 24] & 48 & [B$\times$F, T, 48] \\
\hline
Channel-wise Linear & [B$\times$F, T, 48] & 24 & [B$\times$F, T, 24] \\
\hline
Reshape & [B$\times$F, T, 24] & - & [B, 24, F, T] \\
\hline
 Inplace Deconv2d (6-2) $\times$5 & [B, 48, F, T] & 5$\times$1, (1, 1), 24 & [B, 24, F, T] \\
 \hline
 Inplace Deconv2d (1) (IBN)  & [B, 48, F, T] & 5$\times$1, (1, 1), 24 & [B, 1, F, T] \\
  \hline
 Inplace Deconv2d (1) $\times$4 (ISAM) & [B, 48, F, T] & 5$\times$1, (1, 1), 24 & [B, 24, F, T] \\
 \hline
 \end{tabular}%
 }
 \label{table0}
\end{table}


\setlength{\tabcolsep}{6.45pt}  
\begin{table*}[htbp]
\centering
\scriptsize  
\caption{STOI[\%], ESTOI[\%], PESQ, SI-SDR, WER[\%] and TSOS[\%] for Non-Moving and Moving Datasets.}
\label{table1}
\begin{tabularx}{\textwidth}{c c| c c c c c c|c c c c c c c}
\hline
\multicolumn{2}{c|}{\multirow{2}{*}{\centering \textbf{Model}}} & \multicolumn{6}{c|}{\rule{0pt}{2.0ex}\textbf{Non-Moving Dataset }} & \multicolumn{7}{c}{\rule{0pt}{2.0ex}\textbf{Moving Dataset }} \\
\multicolumn{2}{c|}{} & \textbf{STOI} \makebox[0pt][c]{$\uparrow$} & \textbf{ESTOI} \makebox[0pt][c]{$\uparrow$} & \textbf{PESQ} \makebox[0pt][c]{$\uparrow$} & \textbf{SI-SDR} \makebox[0pt][c]{$\uparrow$} & \textbf{WER} \makebox[0pt][c]{$\downarrow$} & \textbf{TSOS} \makebox[0pt][c]{$\downarrow$} & \textbf{STOI} \makebox[0pt][c]{$\uparrow$} & \textbf{ESTOI} \makebox[0pt][c]{$\uparrow$} & \textbf{PESQ} \makebox[0pt][c]{$\uparrow$} & \textbf{SI-SDR} \makebox[0pt][c]{$\uparrow$} & \textbf{WER} \makebox[0pt][c]{$\downarrow$} & \textbf{TSOS} \makebox[0pt][c]{$\downarrow$} \\
\hline
\multicolumn{2}{c|}{\rule{0pt}{2.0ex} Noisy } & 75.39&  55.03&  2.12& 2.18  & 41.87 & - & 74.93 & 54.71 & 2.12 & 2.09  & 45.94 & - & \\
\hline
\rule{0pt}{2.0ex}
\multirow{3}{*}{\rule{0pt}{3.0ex}\textbf{Non-causal}} & F-Conv-TasNet & 84.18 & 69.74 & 2.46 & 10.12 & 49.46 & 21.74 & 83.99 & 69.35 & 2.48 & 10.05  & 50.63 & 19.63 & \\
& ATT-MVDR & 87.75 & 73.69 & 2.69 & 10.49  &30.43 & 0.86 & 85.47 & 70.57 & 2.63 & 9.78 & 38.01 & \textbf{0.44} & \\
& \textbf{ABIC-MVDR} & \textbf{92.41} & \textbf{81.62} & \textbf{3.05} & \textbf{12.69}  & \textbf{18.63} & \textbf{0.52} & \textbf{91.49} & \textbf{79.98} & \textbf{3.03} & \textbf{12.47}  & \textbf{22.02} & 0.76 & \\
\hline
\rule{0pt}{2.0ex}
\multirow{3}{*}{\rule{0pt}{3.0ex}\textbf{Causal}} & F-Conv-TasNet & 83.09 & 68.03 & 2.41& 9.69  & 50.20 &17.12 & 82.66 & 67.49 & 2.42 & 9.66  & 52.98 & 17.58 & \\
& ATT-MVDR & 86.3 & 71.19 & 2.58 & 9.03  & 32.37 &0.49 & 83.8 & 67.75 & 2.52 & 8.57  & 40.38 & 0.51 & \\
& \textbf{ABIC-MVDR} & \textbf{91.88} & \textbf{80.62} & \textbf{2.99} & \textbf{11.22}  & \textbf{19.82} & \textbf{0.42} & \textbf{90.46} & \textbf{77.99} & \textbf{2.92} & \textbf{11.54}  & \textbf{28.17} & \textbf{0.43} & \\
\hline
\end{tabularx}
\end{table*}

\section{Experiment results and analysis}

\renewcommand{\thefootnote}{\arabic{footnote}}

\subsection{Experimental results}

In this paper, the ATT-MVDR \cite{ochiai2023mask} is employed as the baseline. The output of the first stage in ATT-MVDR is referred to as F-Conv-TasNet \cite{bahmaninezhad2019comprehensive}, which a single-channel model used to estimate the target speech mask, as described in Section \ref{sec:introduction}. Table  \ref{table1} compares our method with baselines across six metrics under different combinations of conditions (moving or non-moving and causal or non-causal).  The causal model uses only past time frames to enable real-time applications, while the non-causal model uses both past and future time frames for better accuracy, as implemented in Section~\ref{sec:ISAM}. Six objective metrics are employed: short-time objective intelligibility (STOI) \cite{taal2011algorithm}, extended short-time objective intelligibility (ESTOI) \cite{jensen2016algorithm}, perceptual evaluation of speech quality (PESQ) \cite{beerends2002perceptual},  scale-invariant signal-to-distortion ratio (SI-SDR) \cite{LeRoux2019}, word error rate (WER), and target speaker over-suppression (TSOS)\cite{EskimezICASSP2022}.
To evaluate the degree of speech distortion, WER and TSOS are employed as key metrics, with lower values indicating less distortion. WER is computed using the Whisper model with its base size\footnote[1]{\url{https://github.com/openai/whisper}}. Table \ref{table3} shows the parameters and Multiply-Accumulate Operations (MACs) for each model, where the MACs are calculated for 1 second audio.


 F-conv-tasnet estimates the time-frequency mask using a neural network, and its WER and TSOS are significantly higher than that of models using MVDR, such as ABIC-MVDR and ATT-MVDR, which shows that one of the advantages of MVDR is that it reduces speech distortion.  As shown in Table \ref{table1} and Table \ref{table3}, the second stage in ATT-MVDR has limited improvement over the first stage, but consumes numerous parameters and computing resources. 
 As discussed in Section \ref{sec:introduction}, one
explanation is the accumulation of errors across the two stages and the lack of consideration for frequency band independence
when calculating attention weights. Our proposed method  outperforms baselines across most metrics under different conditions.  ABIC-MVDR has only 0.35 million trainable parameters, which is 108 times fewer than ATT-MVDR, and requires 4.04 G/s MACs, approximately 1.68 times fewer. This demonstrates that our proposed attention-based beamformer leveraging IGCRN is more efficient in learning spatial information and calculating attention weights for SCM reconstruction compared to ATT-MVDR.

\subsection{Ablation experiments}
We conduct 3 ablation experiments to verify the effectiveness of each module:
\begin{enumerate}
    \item \textbf{ONLINE-MVDR / BLOCK-MVDR}:  
    To validate the facilitation of the ISAM module, this ablation experiment calculates the SCM using online processing [\citenum{higuchi2016robust}, \citenum{8683184}] (ONLINE-MVDR) and blockwise processing \cite{kubo2019mask} (BLOCK-MVDR) based on the ISCM. The forgetting factor is set to 0.995 in ONLINE-MVDR  and the block size is set to 30 in BLOCK-MVDR. 

    \item \textbf{CRN-MVDR}: The IGCRN module in the proposed method is replaced by the CRN\cite{tan2018convolutional} to verify the performance of the IGCRN.
\end{enumerate}

\begin{table}[h]
\centering
\caption{Ablation experiments of ABIC-MVDR on moving datasets.}
\label{table2}
\renewcommand{\arraystretch}{1.2}
\resizebox{\linewidth}{!}{%
\begin{tabular}{c|l|c c c c c c}
\hline
\multicolumn{2}{c|}{\textbf{model}} & \textbf{STOI} \makebox[0pt][c]{$\uparrow$} & \textbf{ESTOI} \makebox[0pt][c]{$\uparrow$} & \textbf{PESQ} \makebox[0pt][c]{$\uparrow$} & \textbf{SI-SDR} \makebox[0pt][c]{$\uparrow$} & \textbf{WER} \makebox[0pt][c]{$\downarrow$} & \textbf{TSOS} \makebox[0pt][c]{$\downarrow$} \\ \hline
\multicolumn{2}{c|}{\textbf{Noisy}} & 74.93 & 54.71 & 2.12 & 2.09 & 45.94 & - \\\hline
\multirow{2}{*}{\textbf{non-causal}} & CRN-MVDR & 90.15 & 77.82 & 2.91 & 11.89 & 26.60 & \textbf{0.62} \\ 
& \textbf{ABIC-MVDR} & \textbf{91.49} & \textbf{79.98} & \textbf{3.03} & \textbf{12.47} & \textbf{22.02} & 0.76 \\ \hline
\multirow{4}{*}{\textbf{causal}} & ONLINE-MVDR & 82.17 & 64.12 & 2.37 & 2.12 & 43.75 & 4.63 \\ 
& BLOCK-MVDR & 82.36 & 63.39 & 2.33 & 3.97 &42.77 & 1.72\\ 
& CRN-MVDR & 89.90 & 77.20 & 2.84 & 11.28 & \textbf{26.89} & 1.00 \\ 
& \textbf{ABIC-MVDR} & \textbf{90.46} & \textbf{77.99} & \textbf{2.92} & \textbf{11.54} & 28.17 & \textbf{0.43} \\ \hline
\end{tabular}%
}
\end{table}

Table \ref{table2} shows the results of the ablation experiments on moving datasets. Compared to ONLINE-MVDR and BLOCK-MVDR, the proposed model demonstrates superior performance. This suggests that the inclusion of the ISAM module enhances the capability of SCM reconstruction, allowing it to allocate weights more accurately over time compared to online or blockwise processing in scenarios with moving sources.
From Table \ref{table2} and \ref{table3}, it can be seen that the proposed method outperforms CRN-MVDR on most metrics, with reduced computation and fewer parameters. This proves that spatial information is effectively extracted through IGCRN for reconstructing SCMs, making it compatible with beamformer.


\begin{table}[h]
\centering
\caption{Comparison of Model Parameters and MACs}
\label{table3}
\begin{tabular}{>{\centering\arraybackslash}p{2.5cm}|>{\centering\arraybackslash}p{2cm}|>{\centering\arraybackslash}p{2cm}} 
\hline
Method & Parameters (M) & MACs (G/s) \\\hline
\rule{0pt}{1.5ex}
F-conv-Tasnet & 8.81 & 2.43 \\
ATT-MVDR      & 37.99 & 6.77 \\
CRN-MVDR     & 19.82 & 6.04 \\
ABIC-MVDR     & 0.35 & 4.04 \\
\hline
\end{tabular}
\end{table}



\section{Conclusion}
This paper proposes a novel attention-based beamformer combining IGCRN for multi-channel speech enhancement.
 The IGCRN leverages the characteristics of the inplace convolution operator and frequency-independent LSTM to learn spatial information. The attention mechanism is applied to allocate temporal weights for reconstructing the speech and noise SCMs accurately. The MVDR beamformer  reduces speech distortion. Experimental results indicate that the proposed method surpasses the baselines across various conditions, with reduced computational complexity and fewer parameters. 

\bibliographystyle{IEEEtran}

\begin{thebibliography}{10}
\providecommand{\url}[1]{#1}
\csname url@samestyle\endcsname
\providecommand{\newblock}{\relax}
\providecommand{\bibinfo}[2]{#2}
\providecommand{\BIBentrySTDinterwordspacing}{\spaceskip=0pt\relax}
\providecommand{\BIBentryALTinterwordstretchfactor}{4}
\providecommand{\BIBentryALTinterwordspacing}{\spaceskip=\fontdimen2\font plus
\BIBentryALTinterwordstretchfactor\fontdimen3\font minus \fontdimen4\font\relax}
\providecommand{\BIBforeignlanguage}[2]{{%
\expandafter\ifx\csname l@#1\endcsname\relax
\typeout{** WARNING: IEEEtran.bst: No hyphenation pattern has been}%
\typeout{** loaded for the language `#1'. Using the pattern for}%
\typeout{** the default language instead.}%
\else
\language=\csname l@#1\endcsname
\fi
#2}}
\providecommand{\BIBdecl}{\relax}
\BIBdecl

\bibitem{benesty2008microphone}
J.~Benesty, J.~Chen, and Y.~Huang, \emph{Microphone Array Signal Processing}.\hskip 1em plus 0.5em minus 0.4em\relax Berlin, Germany: Springer, 2008.

\bibitem{arrays2001signal}
M.~Brandstein and D.~Ward, Eds., \emph{Microphone Arrays: Signal Processing Techniques and Applications}.\hskip 1em plus 0.5em minus 0.4em\relax Berlin, Germany: Springer, 2001.

\bibitem{Benesty2012DifferentialMicrophone}
J.~Benesty and J.~Chen, \emph{Study and Design of Differential Microphone Arrays}.\hskip 1em plus 0.5em minus 0.4em\relax New York, NY, USA: Springer, 2013.

\bibitem{Veen1988Beamforming}
B.~D. Van~Veen and K.~M. Buckley, ``Beamforming: A versatile approach to spatial filtering,'' \emph{IEEE assp magazine}, vol.~5, no.~2, pp. 4--24, 1988.

\bibitem{heymann2016neural}
J.~Heymann, L.~Drude, and R.~Haeb-Umbach, ``Neural network based spectral mask estimation for acoustic beamforming,'' in \emph{2016 IEEE International Conference on Acoustics, Speech and Signal Processing (ICASSP)}.\hskip 1em plus 0.5em minus 0.4em\relax IEEE, 2016, pp. 196--200.

\bibitem{erdogan2016improved}
H.~Erdogan, J.~R. Hershey, S.~Watanabe, M.~I. Mandel, and J.~Le~Roux, ``Improved mvdr beamforming using single-channel mask prediction networks.'' in \emph{Interspeech}, 2016, pp. 1981--1985.

\bibitem{higuchi2016robust}
T.~Higuchi, N.~Ito, T.~Yoshioka, and T.~Nakatani, ``Robust mvdr beamforming using time-frequency masks for online/offline asr in noise,'' in \emph{2016 IEEE International Conference on Acoustics, Speech and Signal Processing (ICASSP)}.\hskip 1em plus 0.5em minus 0.4em\relax IEEE, 2016, pp. 5210--5214.

\bibitem{kubo2019mask}
Y.~Kubo, T.~Nakatani, M.~Delcroix, K.~Kinoshita, and S.~Araki, ``Mask-based mvdr beamformer for noisy multisource environments: Introduction of time-varying spatial covariance model,'' in \emph{ICASSP 2019-2019 IEEE International Conference on Acoustics, Speech and Signal Processing (ICASSP)}.\hskip 1em plus 0.5em minus 0.4em\relax IEEE, 2019, pp. 6855--6859.

\bibitem{8683184}
M.~Togami, ``Simultaneous optimization of forgetting factor and time-frequency mask for block online multi-channel speech enhancement,'' in \emph{ICASSP 2019 - 2019 IEEE International Conference on Acoustics, Speech and Signal Processing (ICASSP)}, 2019, pp. 2702--2706.

\bibitem{ochiai2023mask}
T.~Ochiai, M.~Delcroix, T.~Nakatani, and S.~Araki, ``Mask-based neural beamforming for moving speakers with self-attention-based tracking,'' \emph{IEEE/ACM Transactions on Audio, Speech, and Language Processing}, vol.~31, pp. 835--848, 2023.

\bibitem{vaswani2017attention}
A.~Vaswani, N.~Shazeer, N.~Parmar, J.~Uszkoreit, L.~Jones, A.~N. Gomez, .~Kaiser, and I.~Polosukhin, ``Attention is all you need,'' in \emph{Proceedings of the 31st International Conference on Neural Information Processing Systems}, Long Beach, USA, 2017, pp. 6000--6010.

\bibitem{liu2010wideband}
W.~Liu and S.~Weiss, \emph{Wideband Beamforming: Concepts and Techniques}.\hskip 1em plus 0.5em minus 0.4em\relax Hoboken, NJ, USA: Wiley, 2010.

\bibitem{Liu2021Inplace}
\BIBentryALTinterwordspacing
J.~Liu and X.~Zhang, ``Inplace gated convolutional recurrent neural network for dual-channel speech enhancement,'' in \emph{Interspeech 2021, 22nd Annual Conference of the International Speech Communication Association, Brno, Czechia, 30 August - 3 September 2021}, H.~Hermansky, H.~Cernock{\'{y}}, L.~Burget, L.~Lamel, O.~Scharenborg, and P.~Motl{\'{\i}}cek, Eds.\hskip 1em plus 0.5em minus 0.4em\relax {ISCA}, 2021, pp. 1852--1856. [Online]. Available: \url{https://doi.org/10.21437/Interspeech.2021-899}
\BIBentrySTDinterwordspacing

\bibitem{Tan2019RealtimeSE}
K.~Tan, X.~Zhang, and D.~Wang, ``Real-time speech enhancement using an efficient convolutional recurrent network for dual-microphone mobile phones in close-talk scenarios,'' in \emph{ICASSP 2019-2019 IEEE International Conference on Acoustics, Speech and Signal Processing (ICASSP)}.\hskip 1em plus 0.5em minus 0.4em\relax IEEE, 2019, pp. 5751--5755.

\bibitem{Tan2019LearningCS}
K.~Tan and D.~Wang, ``Learning complex spectral mapping with gated convolutional recurrent networks for monaural speech enhancement,'' \emph{IEEE/ACM Transactions on Audio, Speech, and Language Processing}, vol.~28, pp. 380--390, 2019.

\bibitem{Zhang2022Complex}
C.~Zhang, J.~Liu, and X.~Zhang, ``A complex spectral mapping with inplace convolution recurrent neural networks for acoustic echo cancellation,'' in \emph{ICASSP 2022-2022 IEEE International Conference on Acoustics, Speech and Signal Processing (ICASSP)}.\hskip 1em plus 0.5em minus 0.4em\relax IEEE, 2022, pp. 751--755.

\bibitem{Zhang2022LCSM2}
{C. Zhang}, {J. Liu}, and {X. Zhang}, ``Lcsm: A lightweight complex spectral mapping framework for stereophonic acoustic echo cancellation,'' in \emph{Interspeech}, 2022, pp. 2523--2527.

\bibitem{Ioffe2015BatchNorm}
\BIBentryALTinterwordspacing
S.~Ioffe and C.~Szegedy, ``Batch normalization: Accelerating deep network training by reducing internal covariate shift,'' in \emph{Proceedings of the 32nd International Conference on Machine Learning, {ICML} 2015, Lille, France, 6-11 July 2015}, ser. {JMLR} Workshop and Conference Proceedings, F.~R. Bach and D.~M. Blei, Eds., vol.~37.\hskip 1em plus 0.5em minus 0.4em\relax JMLR.org, 2015, pp. 448--456. [Online]. Available: \url{http://proceedings.mlr.press/v37/ioffe15.html}
\BIBentrySTDinterwordspacing

\bibitem{souden2010optimal}
M.~Souden, J.~Benesty, and S.~Affes, ``On optimal frequency-domain multichannel linear filtering for noise reduction,'' \emph{IEEE Transactions on audio, speech, and language processing}, vol.~18, no.~2, pp. 260--276, 2009.

\bibitem{Paul1992CSR}
\BIBentryALTinterwordspacing
D.~B. Paul and J.~M. Baker, ``The design for the wall street journal-based {CSR} corpus,'' in \emph{The Second International Conference on Spoken Language Processing, {ICSLP} 1992, Banff, Alberta, Canada, October 13-16, 1992}.\hskip 1em plus 0.5em minus 0.4em\relax {ISCA}, 1992, pp. 899--902. [Online]. Available: \url{https://doi.org/10.21437/ICSLP.1992-277}
\BIBentrySTDinterwordspacing

\bibitem{Barker2015CHiME}
J.~Barker, R.~Marxer, E.~Vincent, and S.~Watanabe, ``The third ‘chime’speech separation and recognition challenge: Dataset, task and baselines,'' in \emph{2015 IEEE Workshop on Automatic Speech Recognition and Understanding (ASRU)}.\hskip 1em plus 0.5em minus 0.4em\relax IEEE, 2015, pp. 504--511.

\bibitem{kingma2014adam}
D.~P. Kingma and J.~Ba, ``Adam: A method for stochastic optimization,'' in \emph{3rd International Conference on Learning Representations, ICLR 2015, Conference Track Proceedings}, Y.~Bengio and Y.~LeCun, Eds., San Diego, CA, USA, 2015.

\bibitem{LeRoux2019}
J.~Le~Roux, S.~Wisdom, H.~Erdogan, and J.~R. Hershey, ``Sdr--half-baked or well done?'' in \emph{ICASSP 2019-2019 IEEE International Conference on Acoustics, Speech and Signal Processing (ICASSP)}.\hskip 1em plus 0.5em minus 0.4em\relax IEEE, 2019, pp. 626--630.

\bibitem{bahmaninezhad2019comprehensive}
\BIBentryALTinterwordspacing
F.~Bahmaninezhad, J.~Wu, R.~Gu, S.~Zhang, Y.~Xu, M.~Yu, and D.~Yu, ``A comprehensive study of speech separation: Spectrogram vs waveform separation,'' in \emph{Interspeech 2019, 20th Annual Conference of the International Speech Communication Association, Graz, Austria, 15-19 September 2019}, G.~Kubin and Z.~Kacic, Eds.\hskip 1em plus 0.5em minus 0.4em\relax {ISCA}, 2019, pp. 4574--4578. [Online]. Available: \url{https://doi.org/10.21437/Interspeech.2019-3181}
\BIBentrySTDinterwordspacing

\bibitem{taal2011algorithm}
C.~H. Taal, R.~C. Hendriks, R.~Heusdens, and J.~Jensen, ``An algorithm for intelligibility prediction of time--frequency weighted noisy speech,'' \emph{IEEE Transactions on audio, speech, and language processing}, vol.~19, no.~7, pp. 2125--2136, 2011.

\bibitem{jensen2016algorithm}
J.~Jensen and C.~H. Taal, ``An algorithm for predicting the intelligibility of speech masked by modulated noise maskers,'' \emph{IEEE/ACM Transactions on Audio, Speech, and Language Processing}, vol.~24, no.~11, pp. 2009--2022, 2016.

\bibitem{beerends2002perceptual}
J.~G. Beerends, A.~P. Hekstra, A.~W. Rix, and M.~P. Hollier, ``Perceptual evaluation of speech quality (pesq) the new itu standard for end-to-end speech quality assessment part ii: psychoacoustic model,'' \emph{Journal of the Audio Engineering Society}, vol.~50, no.~10, pp. 765--778, 2002.

\bibitem{EskimezICASSP2022}
S.~E. Eskimez, T.~Yoshioka, H.~Wang, X.~Wang, Z.~Chen, and X.~Huang, ``Personalized speech enhancement: New models and comprehensive evaluation,'' in \emph{ICASSP 2022-2022 IEEE International Conference on Acoustics, Speech and Signal Processing (ICASSP)}.\hskip 1em plus 0.5em minus 0.4em\relax Ieee, 2022, pp. 356--360.

\bibitem{tan2018convolutional}
K.~Tan and D.~Wang, ``A convolutional recurrent neural network for real-time speech enhancement.'' in \emph{Interspeech}, vol. 2018, 2018, pp. 3229--3233.

\end{thebibliography}
\end{document}